\documentclass{aa}  

\bibpunct{(}{)}{;}{a}{}{,} 
\usepackage{natbib}
\usepackage{graphicx}
\usepackage{lscape}
\usepackage{rotating}
\usepackage{subfig}
\usepackage{multirow}
\usepackage{adjustbox}
\usepackage{xcolor,colortbl,tabularx}
\usepackage{graphicx}
\usepackage{caption}
\usepackage{hyperref}
\usepackage[T1]{fontenc}
\usepackage{txfonts}

\newcommand{\rxte}{\textit{RXTE}\xspace}

\begin{document}

   \title{Indications of non-conservative mass-transfer in AMXPs}

   \author{A. Marino\inst{1,2,3}, T. Di Salvo\inst{1}, L. Burderi\inst{4}, A. Sanna\inst{4}, A. Riggio\inst{4}, A. Papitto\inst{5}, M. Del Santo\inst{2},  A. F. Gambino\inst{4}, R. Iaria\inst{1}, S. M. Mazzola\inst{1}
          }

   \institute{Universit\`a degli Studi di
  Palermo, Dipartimento di Fisica e Chimica, via Archirafi 36 - 90123 Palermo, Italy
              \and
             INAF/IASF Palermo, via Ugo La Malfa 153, I-90146 - Palermo, Italy 
             \and 
             IRAP, Universit\`e de Toulouse, CNRS, UPS, CNES, Toulouse, France
             \and
             Universit\`a degli Studi di
  Cagliari, Dipartimento di Fisica, SP Monserrato-Sestu km 0.7, I-09042 Monserrato, Italy
  			\and
            INAF, Osservatorio Astronomico di Roma, Via di Frascati 33, I-00044, Monteporzio Catone (Roma), Italy\\}

 
  \abstract
   {Since the discovery of the first Accreting Millisecond X-ray Pulsar SAX J1808.4-3658 in 1998, the family of these sources kept growing on. Currently, it counts 22 members. All AMXPs are transients with usually very long quiescence periods, implying that mass accretion rate in these systems is quite low and not constant. Moreover, for at least three sources, a non-conservative evolution was also proposed.}
   {Our purpose is to study the long term averaged mass-accretion rates in all the Accreting Millisecond X-ray Pulsars discovered so far, to investigate a non-conservative mass-transfer scenario.}
   {We calculated the expected mass-transfer rate under the hypothesis of a conservative evolution based on their orbital periods and on the (minimum) mass of the secondary (as derived from the mass function), driven by gravitational radiation and/or magnetic braking. Using this theoretical mass-transfer, we determined the expected accretion luminosity of the systems.  Thus, we achieved the lower limit to the distance of the sources by comparing the computed theoretical luminosity and the observed flux averaged over a time period of 20 years. Then, the lower limit to the distance of the sources has been compared to the value of the distance reported in literature to evaluate how reasonable is the hypothesis of a conservative mass-transfer.}
   {Based on a sample of 18 sources, we found strong evidences of a non-conservative mass-transfer for five sources, for which the estimated distance lower limits are higher than their known distances. We also report hints for mass outflows in other six sources. The discrepancy can be fixed under the hypothesis of a non-conservative mass-transfer in which a fraction of the mass transferred onto the compact object is swept away from the system, likely due to the (rotating magnetic dipole) radiation pressure of the pulsar.}
   {}
              

  \authorrunning{Marino, A.}
  \maketitle
  \email{alessio.marino@unipa.it}
%

\section{Introduction}
\label{sec:intro}
Accreting Millisecond X-ray Pulsars (AMXPs) are Low Mass X-ray Binaries (LMXBs) hosting quickly rotating pulsars with periods of a few milliseconds accreting matter from their low-mass companion stars. Their discovery was a key chapter for modern Astrophysics, since it represented a first step in the confirmation of the so-called \emph{recycling scenario} \citep[see, e.g.][]{Bhattacharya1991}. The aim of this theory was to explain the peculiar case of Millisecond Radio Pulsars (MSPs), which exhibited the weak magnetic field ($\sim 10^8$-$10^9$ G) of old neutron stars and at the same time a spin frequency so high to be incompatible with the scenario of an old isolated neutron star. According to the recycling scenario, MSPs are the leftovers of LMXBs hosting a neutron star (NS) once mass accretion has ended. 
In these systems, when mass transfer is active, the transferred matter, due to its own specific angular momentum, does not fall in a straight line onto the compact object, but it rather spirals around it forming an accretion disc. This disk is expected to be truncated close to the Neutron Star, due to the weak magnetic field of the old compact object. If the Keplerian spin frequency of matter overcomes the NS spin frequency in correspondence of the inner radius of the accretion disk, the NS is spun-up to very short periods, in principle up to the mass shedding period \citep[below $\sim$ 1 ms, depending on the equation of state of ultra-dense matter; for a review, see, e.g., ][]{Ozel2016}. LMXBs and MSPs were two distinct classes of sources, theoretically connected by the recycling scenario, until 1998, when the NASA X-ray Observatory Rossi X-ray Timing Explorer (\emph{RXTE}) discovered the first LMXB exhibiting X-ray coherent pulsations, i.e. SAX J1808.4-3658 \citep{Wijnands1998}. \\ Since then, other 21 AMXPs have been discovered with spin periods ranging between 1.7 and 9.5 ms \citep[see, e.g. ][]{Patruno2012,Campana2018}, the most recent one discovered by \emph{NICER} \citep{Strohmayer2018,Sanna2018}. Indeed, the direct confirmation of the recycling scenario dates back to 2013: \emph{XMM-Newton} observed a source previously classified as a radio millisecond pulsar, i.e. IGR J18245-2452, that behaved as an AMXP during one X-ray active state \citep{Papitto2013}. The source showed swings between X-ray, accretion-powered pulsations to radio, rotation-powered pulsations over short time-scales (less than a couple of weeks). Along with PSR J1023+0038\footnote{Note that it has been recently proposed that PSR J1023+0038 might not be in an accretion-powered pulsar phase even when the X-ray pulsations are clearly detected \citep{Papitto2019arx}} \citep{Archibald2009,Archibald2013} and XSS J12270-4859 \citep{DiMartino2013,Bassa2014,Papitto2015}, IGR J18245-2452 is part of a sub-class of MSPs dubbed transitional millisecond pulsars. These sources are the living proof that radio MSPs, AMXPs and LMXBs may be considered different seasons of the lifetime of a low-mass binary system hosting a neutron star (see, for a review \citealt{Campana2018}).  \\ AMXPs are usually found in compact binary systems, as witnessed by their relatively short orbital periods (with a few exceptions, the most relevant one is Aql X-1, which has an orbital period of 19 h, \citealt{Chevalier1991,Welsh2000}), from $\sim$ 40 min to a few hours, and therefore they probably host very low-mass donor stars, with $M \leq 0.2 M_\odot$. All AMXPs known so far are X-ray transients; some of them show outbursts every few years (such as SAX J1808.4-3658, which goes in outburst every $\sim$ 2.5 years) while others have been observed in outburst only once since their discovery, e.g. XTE J0929-314 and XTE J1807-294 \citep{Galloway2002,Riggio2008}.  The duration of the outbursts can also be quite long, as in the case of HETE J1900.1-2455 \citep[in outburst for $\sim$ 10 years,][]{Simon2018} and MAXI J0911-655 \citep[which is in an ongoing outburst since 2016,][]{Sanna2017b}. In order to explain the strong orbital expansion of the AMXPs SAX J1808.4-3658 \citep{DiSalvo2008, Burderi2009a,Sanna2017} and SAX J1748.9-2021 \citep{Sanna2016} it was proposed that only a fraction of the mass transferred from the secondary is effectively accreted onto the neutron star \citep[see also][]{Tailo2018}. Matter ejections have been also suggested to explain the low, average, mass-transfer rate derived for XTE J1814-338 \citep{Van2018}.
This scenario has been recently confirmed with an independent argument by \citet{Marino2017} for XTE J0929-314, for which a non-conservative mass-transfer has been invoked to explain the discrepancy between the observed averaged luminosity and the expected luminosity, estimated on the basis of a conservative evolutionary model driven by Gravitational Radiation (GR) \citep{Verbunt1995}. \\
In this paper we use the same argument as for XTE J0929-314 and we apply it to almost the complete sample of AMXPs, with the aim of exploring how common (or uncommon) non-conservative mass-transfer is among such sources. In Table \ref{tab:amxps} we present a summary of the main properties of these sources, i.e. spin period $P_{ms}$, orbital period $P_{orb}$, mass function $f_{X}$, the minimum mass for the companion star $M_{2,min}$, the distance and the year when pulsations were observed for the first time. The table is an updated version of Table 1 in \cite{Campana2018}. The minimum mass is estimated from the mass function $f_X$ fixing the inclination of the system to 90$^\circ$.

\begin{table*}
\centering
\caption{\textbf{List of the AMXPs discovered so far and their main properties}}
\begin{tabular}{l  l  l  l  l  l  l  l  l}
\hline 
\hline
& \textbf{P$_{orb}$} & \textbf{P$_s$} & \textbf{f$_X$} & \textbf{M$_{2,min}$} & \textbf{Year of}	& Ref.\\
{\textbf{Source}} & {(hr)} &  (ms) & (M$_{\odot}$) & (M$_{\odot}$) & \textbf{discovery} & \\
\hline
Aql X-1 & 18.95 & 1.7 & 1.4$\times$10$^{-2}$ & 0.56 & 2008 & [6], [14]\\
IGR J18245-2452 & 11.03 & 3.9 & 2.3$\times$10$^{-3}$ & 0.17 & 2013 & [17]\\
Swift J1749.4–2807 & 8.82 & 1.9 &  5.5$\times$10$^{-2}$ & 0.59 & 2010 & [3]\\
IGR J17591-2342 & 8.80 & 1.9 & 1.5$\times$10$^{-2}$ & 0.42 & 2018 & [23]\\  
SAX J1748.9–2021 & 8.77 & 2.3 & 4.8$\times$10$^{-3}$ & 0.10 & 2008 & [1]\\
XSS J12270–4859 & 6.91 & 1.7 & 3.9$\times$10$^{-3}$ & 0.27 & 2015 & [18]\\
PSR J1023+0038 & 4.75 & 1.7 & 1.1$\times$10$^{-3}$ & 0.20 & 2009 & [4],  [7]\\
XTE J1814–338 & 4.27 & 3.2 & 2.0$\times$10$^{-3}$ & 0.17 & 2003 & [13]\\
IGR J17498–2921 & 3.84 & 2.5 & 2.0$\times$10$^{-3}$ & 0.17 & 2011 & [16]\\
IGR J17511–3057 & 3.47 & 4.1 & 1.1$\times$10$^{-3}$ & 0.13 & 2010 & [15] \\
IGR J00291+5934 & 2.46 & 1.7 & 2.8$\times$10$^{-3}$ & 0.039 & 2005 & [9] \\
SAX J1808.4–3658 & 2.01 & 2.5 & 3.8$\times$10$^{-5}$ & 0.043 & 1998 & [25] \\
IGR J1737.9-3747 & 1.88 & 2.1 & 8.5$\times$10$^{-5}$ & 0.056 & 2018 & [21] \\
HETE J1900.1-2455 & 1.39 & 2.7 & 2.0$\times$10$^{-6}$ & 0.016 & 2006 & [10]\\
NGC6440 X–2 & 0.95 & 4.8 & 1.6$\times$10$^{-7}$ & 0.00067 & 2010 & [2]\\
Swift J1756.9–2508 & 0.91 & 5.5 & 1.6$\times$10$^{-7}$ & 0.0070 & 2007 & [11]\\
IGR J16597–3704 & 0.77 & 9.5 & 1.2$\times$10$^{-7}$ & 0.0060 & 2017 & [20]\\
MAXI J0911–655 & 0.74 & 2.9 & 6.2$\times$10$^{-6}$ & 0.024 & 2016 & [19]\\
XTE J0929–314 & 0.73 & 5.4 & 2.9$\times$10$^{-7}$ & 0.0083 & 2002 & [8]\\
XTE J1751–305 & 0.71 & 2.3 & 1.3$\times$10$^{-6}$ & 0.014 & 2002 & [12] \\
XTE J1807–294 & 0.67 & 5.3 & 1.5$\times$10$^{-7}$ & 0.0066 & 2003 & [5]\\
IGR J17062-6143 & 0.64 & 6.1 & 9.1$\times$10$^{-8}$ & 0.00060 & 2017 & [24] \\
\hline
\hline
\end{tabular}
\caption*{In the table P$_{orb}$ is the orbital period, P$_s$ is the spin period of the pulsar, f$_X$ is the mass function and M$_{2,min}$ is the minimum mass for the companion, evaluated from f$_X$ for an inclination of 90$^\circ$. References: [1]=\cite{Altamirano2008_1748}, [2]=\cite{Altamirano2010_ngc}, [3]=\cite{Altamirano2011_1749}, [4]=\cite{Archibald2009}, [5]=\cite{Campana2003_1807}, [6]=\cite{Casella2008}, [7]=\cite{Cotizelati2014}, [8]=\cite{Galloway2002}, [9]=\cite{Galloway2005}, [10]=\cite{Kaaret2006}, [11]=\cite{Krimm2007}, [12]=\cite{Markwardt2002}, [13]=\cite{Markwardt2003b}, [14]=\cite{MataSanchez2017}, [15]=\cite{Papitto2010_17511}, [16]=\cite{Papitto2011}, [17]=\cite{Papitto2013}, [18]=\cite{Roy2015}, [19]=\cite{Sanna2017b}, [20]=\cite{Sanna2018_16597}, [21]=\cite{Sanna2018_1737}, [22]=\cite{Sanna2018_J1756}, [23]=\cite{Sanna2018}, [24]=\cite{Strohmayer2018_j17062}, [25]=\cite{Wijnands1998}}
\label{tab:amxps}
\end{table*}

\section{The method}
\label{sec:method}
The method applied here is based on the comparison between the expected average luminosity, $L_{exp}$, expected in the case of an orbital evolution induced by conservative mass-transfer, and the observed averaged X-ray luminosity. This comparison gives an estimation of the lower limit to the source distance, which is used as a test of how reliable the conservative hypothesis is. The same strategy has been already applied to several non-burster AMXPs in order to constrain their distances \citep{Galloway2006} and to the AMXP XTE J0929-314, with the aim of proving the unlikeliness of a conservative mass-transfer in this system \citep{Marino2017}.\\
\subsection{Expected Luminosity} \label{subsec:ltheor}
The expected luminosity of a LMXB might be simply expressed as $L_{exp}= -\frac{GM_1\dot{M_2}}{R}$, where $M_1$ is the mass of the accretor while $\dot{M}_2$ is the (intrinsically negative in the considered case) mass-transfer rate of the donor, in the hypothesis that the whole mass transferred from the secondary is accreted onto the NS.  Assumptions can be made about $M_1$ and $R$ according to the standard values associated to neutron stars and/or the latest results in the search for the NS mass and radius. In order to have a lower limit for the expected luminosity we should choose a lower limit for both the masses and an upper limit for the radius. Here we assumed the mass of the NS equal to 1.4 $M_\odot$, which is low enough to be considered a reasonable lower limit giving that the current record-holder for the least massive NS is PSR J0453+1559, with a mass of 1.174$\pm$0.001 $M_\odot$ \citep{Martinez2015}. Furthermore, according to \citet{Ozel2012}, the distribution of the NS masses is double-peaked, with two maxima corresponing to  $1.28 \ M_\odot$ (with a dispersion of $0.24 \ M_\odot$) and $ 1.54 \pm 0.23 \ M_\odot$ (with a dispersion of $0.20 \ M_\odot$), for NSs in non-recycled eclipsing high-mass binaries and for slow pulsars or NSs in recycled systems, which have experienced several episodes of accretion, respectively. The NSs inhabiting the sources analyzed in the present work belong to the second family, therefore a $1.4$ M$_\odot$ seems like a reasonable choice. Concerning the radius, we chose $12$ km as an upper limit, based on the 9.9-11.2 km range reported by \cite{Ozel2016}. \\ The minimum donor mass for each system was determined on the basis of its mass function $f$, reported in Table \ref{tab:amxps}. Considering the lack of eclipses and/or dips ever observed for all the systems in the class with the exception of one - Swift J1749.4–2807 \citep{Altamirano2011_1749} - we exclude inclination angles $i > 60^\circ$, and we estimated the lower limit for the secondary accordingly. Coherently, for the only eclipsing AMXP discovered so far, this lower limit was estimated fixing the inclination to 90$^\circ$. \\
An estimate of $\dot{M}_2$ requires the introduction of a theoretical orbital evolution model. In the following, we adopt a simple and general orbital evolution model obtained combining the Kepler's third law with the condition that the neutron star is accreting matter from the companion via Roche Lobe overflow. The latter can be expressed as: 
\begin{equation}
\frac{\dot{R}_2}{R_2}=\frac{\dot{R}_{L2}}{R_{L2}} \ ,
\label{eq:condition} 
\end{equation} 
where $R_2$ and $R_{L2}$ are the radii of the secondary and of its Roche Lobe, respectively. Equation \ref{eq:condition} guarantees that during the whole evolution of the system the secondary star fills the Roche Lobe. We assume then a mass-radius relation $R_2 \propto M^n_2$, with $M_2$ the mass of the secondary, and the \citet{Paczynski1971} approximation for the Roche lobe radius $R_{L2}=2/3^{4/3}[q/(1+q)]^{1/3}a$ (valid for $q=M_2/M_1 \leq 0.8$), where $a$ is the orbital separation. We are also taking into account two possible mechanisms of angular momentum loss that, in turn, drive the mass-transfer process from the donor star: gravitational radiation (GR) and magnetic braking (MB). According to this model, the mass-transfer rate $\dot{m}_{-9}$, in units of $10^{-9} M_\odot  yr^{-1}$ can be expressed as:
\begin{equation}
\dot{m}_{-9}=4.03 \ m^{8/3}_1P^{-8/3}_{2h}\left[\frac{q^2}{\left(2g(\alpha,\beta,q)-\frac{1}{3}+n\right)(1+q)^{1/3}}\right]\times[1+T_{MB}]
\label{eq:mdot}
\end{equation}

where:
\begin{equation}
g(\alpha, \beta, q)= 1-\beta q-\frac{1-\beta}{1+q}\left(\frac{q}{3}+\alpha\right) \ ,
\label{eq:gterm}
\end{equation}
\citep[see][]{DiSalvo2008,Burderi2009a,Burderi2010} where $m_1$ is the mass of the primary star in units of $M_\odot$, $P_{2h}$ is the orbital period in units of 2h, $n$ is the index of the mass-radius relation, $\beta$ is the fraction of mass lost by the donor and accreted onto the NS, $\alpha$ is the specific angular momentum carried by the matter ejected from the system in the case of a non-conservative mass-transfer and $T_{MB}$ is the strength of the torque associated to MB. The parameter $T_{MB}$ can be parametrised in units of the GR torque as in \cite{Burderi2010}\footnote{In the cited equation two typos were present, in particular the constant was wrong and the factor $(1+q)^{2/3}$ was missing, therefore we report a corrected version of the same formula.}, in line with \citet{Verbunt1981,Verbunt1994,Tauris2001}, as:
\begin{equation}
T_{MB}=8.4(k^2)_{0.1}f^{-2}m^{-1}_1 P^2_{2h}q^{1/3}(1+q)^{2/3}
\end{equation}
where $k$ is the gyration radius of the secondary star and $f$ is a dimensionless parameter which assumes a value of 0.79 or 1.78 \citep{Skumanich1972,Smith1979}.
In order to estimate the expected mass-transfer rate in the conservative case we take $\beta=1$. \\ The value of $\dot{m}$, given by Equation \ref{eq:mdot}, is highly dependent on the mass of the secondary. On the one hand, because of the assumed mass-radius relation, the value of $n$ changes according to $M_2$, on the other hand the MB is considered to become negligible ($T_{MB}$=0) in Ultra-Compact X-ray Binaries for fully convective stars with $M_2 \leq 0.3 M_\odot$ \citep{Nelson2003}. However, the latter assumption has been questioned at least in one case. Indeed magnetic braking has been included to describe the evolutionary path of the AMXP SAX J1808.4-3658 during its whole history \citep{Wang2013,Tailo2018}, although its companion star is likely a brown dwarf with a mass $\leq 0.1$ $M_\odot$. For uniformity we considered both models, GR and GR+MB, to describe the mass-transfer in each system, fixing $f$ to 
1.78 in order to have a lower limit on $T_{MB}$. In the following we will refer with $L_{exp,GR}$ to the expected luminosity in the GR-only case, and with $L_{exp,GR+MB}$ to the luminosity in the MB-included case. \\
Another assumption concerns the value attributed to $n$. In the case of a low-mass secondary star with $M_2 \leq 0.2 M_\odot$ we chose $n=-1/3$, the proper mass-radius index for fully convective or degenerate stars \citep{Nelson2003}. On the other hand, for masses $M_2 \geq 0.40$ M$_\odot$ we assumed $n=1$ to be a valid approximation  \citep{Chabrier2000}.
Finally we used the value of $k$ computed for low-mass ($M_2 \sim 0.6$ M$_\odot$) stars \citep{Claret1990}, i.e. $k=0.378$. This value is expected to decrease for decreasing masses, therefore the estimated values for $T_{MB}$ in the sources with secondary stars with masses $M_2 << 0.6 \ M_\odot$ have to be considered as upper limits. Note, however, that the MB strength becomes negligible for small values of the orbital separation. \\
\subsection{Observed Luminosity} \label{subsec:lobs}
To estimate the observed average luminosity we considered the energy released by the system during the outburst phase(s) as a good approximation of the whole amount of energy ever emitted by the source, ignoring both the energy released during type-I X-ray bursts and the energy radiated when the source was quiescent. Even if type-I X-ray bursts are among the most energetic displays of LMXBs activity, they usually last for less than an hour; therefore the amount of energy released is negligible if averaged over a baseline of several years. On the other hand, quiescence luminosity is likely at least 3 orders of magnitudes lower than the luminosity in outburst, a ratio which is significantly lower than the ratio between the duration of the outburst and the recurrence time between subsequent outbursts observed so far for AMXPs. Furthermore, accretion is considered to be almost shut off during quiescence \citep[although residual accretion processes might still be ongoing in some quiescent LMXBs, see e.g. ][]{Wijnands2015}, it is therefore reasonable to neglect the energy emitted during this phase. 

A complication to this assumption could be represented by the possibility of "very faint" activity during quiescence, i.e. showing episodes of accretion at low luminosity (around 5$\times$10$^{33}$). Such outbursts would be too faint to be detected from any all-sky monitor and could be seen only with pointed observations. This type of peculiar behavior has been observed in transitional millisecond pulsars, lasting also for several years \citep{Papitto2013,Linares2014,Patruno2014}, although also in this case the contribution of low-level
accretion is not clear yet \citep[see, e.g.][]{Ambrosino2017}. However, the correction to the calculated total energy output when considering these episodes would be significant, i.e. of the order of 10-20\%, only in the case of decades of continuous accretion at $\sim 10^{33}$ erg/s. We then decided to neglect the energy emitted outside the outbursts. \\
Keeping this caveat in mind, the energy released during an outburst was then roughly estimated by calculating the area subtended by the light curve of the system in outburst, which is the observed fluence $f,$ i.e. the total energy per unit area emitted during the outburst as received by the detector. To convert this fluence in the total energy radiated during the outburst, we have to multiply it by a factor of $4\pi d^2$. We consider the distance as a free parameter for the moment and keep this total energy in the form $f\times4\pi d^2$. In order to find the observed luminosity, i.e. F$_{av}\times$4$\pi d^2$, where F$_{av}$ is the average observed flux, one has to average the energy emitted during the outbursts, $f\times4\pi d^2$, all along the period $T$ the source was monitored by our observatories. \\
We considered $T$ equal to 20 years, because since 1996 the X-ray Sky was continuously monitored and any previous outburst would have been therefore recorded. In fact, in 1996, the ASM onboard RXTE and the Wide Field Cameras (WFC) onboard BeppoSAX started a continuous monitoring of the X- ray sky. This is today continued by MAXI onboard the International Space Station, the Swift/BAT (Burst Alert Telescope) hard X-ray monitor, INTEGRAL, and the Gamma-ray Burst Monitor (GBM) on board Fermi. A more conservative choice for $T$ might have been the time elapsed since the discovery of the source, but it would leave out the most recently discovered AMXPs. However, in Section \ref{sec:disc} a comparison between the results obtained with the 20 years-interval and the results obtained with the latter strategy will be presented. \\
Finally, under the hypothesis of a conservative mass-transfer, we can solve the equation F$_{av}\times$4$\pi$d$^2=L_{exp}$ in order to find the distance $d$ to the source; the discrepancy between our value of $d$ and the value reported in literature is then used to quantify the unlikeliness of a conservative mass-transfer. Indeed, in the case of a non-conservative mass-transfer, the mass accretion rate would be lower than the mass transfer rate from the companion star, and this would result in a lower $L_{exp}$ and then in lower values for $d$, fixing in this way the discrepancy. 

\section{Data Analysis}
We analyzed all the sources listed in Table \ref{tab:amxps}, with the exclusion of three systems: Aql X-1, XSS J12270-4859, PSR J1023+0038. Aql X-1 is a peculiar source, given its relatively long orbital period with respect to every other AMXP and the fact that showed coherent X-ray pulsations only during a single $\sim$150-s long data segment \citep[out of $\sim$1~Ms of \rxte\ data][]{Casella2008}. Furthermore it displays a $\sim$ 70 days outburst per year \citep{Ootes2018}, therefore it is expected to have a high averaged mass-accretion rate, likely compatible with a conservative mass-transfer scenario. On the other hand the transitional MSPs XSS J12270-4859 and PSR J1023+0038 have been observed in X-rays only in short, sub-luminous outbursts or in flaring activity at low luminosity, i.e. L$_X \lesssim 10^{34}$ erg/s, under the detectability threshold of any X-ray all-sky monitor \citep{Papitto2015, Cotizelati2014}. Quantifying the energy emitted during these episodes is not an easy task and, at least in the case of PSR J1023+0038, the emitted X-rays flux may not be related to accretion at all, as recently suggested by \cite{Papitto2019arx}. Furthermore, a non-conservative mass-transfer scenario has been already hypothesized to explain the peculiar phenomenology of these systems \citep[as suggested by][]{Heinke2015}. 
In the following we discuss the analysis carried out for the other nineteen sources. In this paper, the results for XTE J0929-314 are also included for sake of completeness, even if its analysis has been extensively discussed by \cite{Marino2017}. \\
For each source, we followed the methodology explained in subsection \ref{subsec:ltheor} in order to estimate the expected luminosity in both the case GR-only and MB-included. The results are reported in Table \ref{tab:results}. For the moment, we used only the minimum mass $M_2$ for the companion, corresponding to the high inclination case, to have a lower limit on the distance of the source.

The estimation of the average observed flux F$_{av}$ requires a study of the observational history of the sources in the last 20 years. We searched in the literature for published light curves of each outburst displayed by each AMXP in the sample. We analysed these light curves numerically, i.e. approximating their shape with a piecewise linear function and then calculating the area subtended by the function. In the analysed light curves, usually, the count rate of the instrument or the flux in a limited energy band is reported rather than the bolometric unabsorbed flux. We used the \emph{HEASARC} Count Rate Simulator \textsc{WebPIMMS} to convert count-rates and/or fluxes into unabsorbed 0.1-300 keV fluxes. In \textsc{WebPIMMS} we described the spectrum of each source using a power-law, with the values of N$_H$ and $\Gamma$ reported in literature for that source. If the flux of the source was reported in Crab units, we assumed the Crab $\Gamma$ index, i.e. $\Gamma=2.1$, to convert the reported flux in bolometric flux. This procedure was not followed for IGR J00291+5934 and XTE J1751-305, whose outburst properties, including peak bolometric flux and outburst durations, were already reported in literature \citep[respectively by ][]{DeFalco2017, Riggio2011}. The spectral parameters of two sources during their outbursts were not available in literature, which are XTE J1814+338 and NGC 6440 X-2. While for the latter we described each of its short, subsequent outbursts with the same parameters reported by \cite{Heinke2010} for the first outburst, for XTE J1814+338 a rough spectral analysis of the source was performed using the \emph{RXTE} standard products. Further details about this analysis are reported in the Appendix. Finally, we followed a different procedure for IGR J17062-6143. Since the source has been persistently in a faint outburst from 2006 \citep{Churazov2007}, the year of its discovery, at a 2-20 keV luminosity of 5.8-7$\times$10$^{35}$ erg/s \citep[see ][ and references therein]{Strohmayer2018_j17062}, we associated to the source a constant luminosity of 6.5$\times$10$^{35}$ erg/s in the 2-20 keV band. Using the known distance of the source, i.e. 7.3 kpc \citep{Keek2017_j17062}, we estimated the flux corresponding to this arbitrary, but compatible with the observations, luminosity.  Finally we extracted the bolometric unabsorbed flux by means of \textsc{WebPIMMS} and calculated $F_{av}\times4\pi d^2$ accordingly.\\
The light curves of some outbursts of a few different sources have never been published in literature. This was the case for IGR J17591-2342, SAX J1748.9-2021, SAX J1808.4-3658, IGR J1737.9-3747 and Swift J1756.9-2508 (see Table \ref{tab:outbursts}). For these cases, we used the archival data of all-sky monitors such as \emph{ASM} onboard \emph{RXTE}, \emph{BAT} onboard \emph{Swift} and \emph{MAXI}, in order to reproduce the light curves of their outbursts\footnote{The public data used here were downloaded from the online archives: \url{https://heasarc.gsfc.nasa.gov/docs/xte/ASM/sources.html}, \url{https://swift.gsfc.nasa.gov/results/transients/} and \url{http://maxi.riken.jp/top/index.html}.}. \\
The outburst properties (duration, peak bolometric flux, etc..) of the AMXPs are summarized in Table \ref{tab:outbursts}. The phenomenology of AMXPs is far from being homogeneous; while most of the sources in the class (10 over 19, including XTE J0929-314) have undergone only one outburst in at least 20 years, almost the other half has shown multiple outbursts. An interesting case is represented by NGC 6440 X-2, which between 2009 and 2011 showed frequent but faint and short (around 3 days each) outbursts \citep{Patruno2013}.

\begin{sidewaystable*}
\centering
\caption{\textbf{Outburst properties.}}
\begin{adjustbox}{max width=\textwidth}
\begin{tabular}{l  l  l  l  l  l l}
\hline 
\hline
& \textbf{\# of} &  & \textbf{Duration} & \textbf{Bolometric peak flux} & \textbf{References} & \textbf{References}\\
{\textbf{Source}} & {\bf outbursts}  & \textbf{Year} & (days) & ($\times 10^{-8}$ erg cm$^{-2}$ s$^{-1}$) & \textbf{Light curves} & \textbf{Spectral Parameters}\\
\hline
IGR J18245-2452 & 1 & 2014 & 23 & 1.15 & [4] & [4]\\
Swift J1749.4–2807 & 1 & 2010 & 11 & 1.52 & [7] & [7]\\
IGR J17591-2342 & 1 & 2018 & 50 & 1.50 & [b] & [32]\\  
SAX J1748.9–2021 & 6 & 1998, 2001, 2005, 2010 & 10, 53, 52, 53 & 0.65, 1.30, 2.10, 4.20, & [1], [1], [1], [m], & [1], [1], [1], [25], \\
& & 2015, 2017 & 58, 13 & 2.09, 1.73, 0.10 & [m], [28] & [27], [28] \\
XTE J1814–338 & 1 & 2003 & 53 & 0.88 & [11] & * \\
IGR J17498–2921 & 1 & 2011 & 37 & 2.00 & [8] & [8]\\
IGR J17511–3057 & 2 & 2010, 2015 & 30, 25 & 0.40, 0.44 & [20], [23] & [20], [23]\\
IGR J00291+5934 & 4 & 2004, 2008, 2008, 2015 & 14, 9, 15, 25 & 0.29, 0.15, 0.11, 0.35 & [5] (all) & [5] (all) \\
SAX J1808.4–3658 & 7 & 1998, 2000, 2002, 2005 & 15, 14, 42, 28, & 0.72, 0.12, 0.92, 0.63, & [11], [11], [11], [11], & [12], [38], [18], [10],\\
& &  2008, 2011, 2015 & 21, 16, 25 & 3.31, 2.24, 0.94 &  [3], [m], [19] & [3], [21], [6] \\
IGR J1737.9-3747 & 3 & 2004, 2008, 2018 & 50, 15, 10 & 0.330, 0.690, 0.014 & [18], [18], [b] & [10], [10], [33]  \\
HETE J1900.1-2455 & 1 & 2005 & 3784 & 2.40 & [36] & [21] \\
NGC6440 X–2 & 10 & 2009, 2009, 2010, 2010, & 3 (all) & 0.088, 0.35, 0.053, 0.088 & [15] (all) & [13] \\
& & & 2010, 2010, 2011, 2011, & 0.035, 0.079, 0.079, 0.026, \\
& & & 2011, 2011 & 0.079, 0.026 \\
Swift J1756.9–2508 & 3 & 2007, 2009, 2018 & 7, 15, 7 & 0.20, 0.26, 0.10 & [15], [a], [2] & [15], [24], [34] \\
IGR J16597–3704 & 1 & 2018 & 17 & 1.80 & [b] & [34] \\
MAXI J0911–655 & 1 & 2016$^\dagger$ & 910 & 0.13 & [31],[x] & [31] \\
XTE J1751–305 & 4 & 2002, 2005, 2007, 2009 & 7, 3, 3, 3 & 0.81, 0.11, 0.14, 0.22 & [29] (all) & [29] (all) \\
XTE J1807–294 & 1 & 2004 & 100 & 0.49 & [7] & [7] \\
IGR J17062-6143 & 1 & 2006$^\dagger$ & 4380 & 0.025 & [37] & [14] \\
\hline
\hline
\end{tabular}
\end{adjustbox}
\caption*{List of the AMXPs discovered so far and their main properties. The bolometric peak flux is the unabsorbed flux in the range 0.1-300 keV and an intrinsic standard 10\% error was attributed to each value. The unpublished light curves were extracted from the public archives of: [a]=\emph{All Sky Monitor} (ASM) onboard \rxte, [b]=\emph{BAT} onboard Swift, [m]=\emph{MAXI} and [x]={\it XRT} onboard \emph{Swift}. The references are divided in refereed publications where the light curve of the relative outburst is displayed and publications where the spectral parameters of the source during the outburst were reported. $^\dagger$: the outburst is still ongoing at the time of writing.
References:
[1]=\cite{Altamirano2008_1748}, [2]=\cite{Bult2018}, [3]=\cite{Cackett2009} [4]=\cite{DeFalco2017_1824}, [5]=\cite{DeFalco2017}, [6]=\cite{DiSalvo2019}, [7]=\cite{Falanga2005_1807}, [8]=\cite{Falanga2012}, [9]=\cite{Ferrigno2011}, [10]=\cite{Galloway2005_1808} [11]=\cite{Hartman2008}, [12]=\cite{Heindl1998}, [13]=\cite{Heinke2010}, [14]=\cite{Keek2017_j17062}, [15]=\cite{Krimm2007}, [16]=\cite{Linares2008}, [17]=\cite{Markwardt2002_1808}, [18]=\cite{Markwardt2008_Atel_1737}, [19]=\cite{Papitto2007}, [20]=\cite{Papitto2010_17511}, [21]=\cite{Papitto2011_1808}, [22]=\cite{Papitto2013_hete}, [23]=\cite{Papitto2016_17511}, [24]=\cite{Patruno2009_J1756}, [25]=\cite{Patruno2010_1748}, [26]=\cite{Patruno2013}, [27]=\cite{Pintore2016} [28]=\cite{Pintore2017}, [29]=\cite{Riggio2011}, [30]=\cite{Sanna2016}, [31]=\cite{Sanna2017b}, [32]=\cite{Sanna2018}, [33]=\cite{Sanna2018_1737}, [34]=\cite{Sanna2018_J1756}, [35]=\cite{Sanna2018_16597}, [36]=\cite{Simon2018}, [37]=\cite{Strohmayer2018_j17062}, [38]=\cite{Wijnands2003_1808}}
\label{tab:outbursts}
\end{sidewaystable*}

The analysis of each outburst, by means of the values reported in Table \ref{tab:outbursts}, led to the estimation of the fluence of the source. Following the recipe of subsection \ref{subsec:lobs}, we calculated the observed luminosity of the source. By equating $L_{exp, GR}$ ($L_{exp, GR+MB}$) and $F_{av}\times 4\pi d^2$ we find the distance of the source under the hypothesis of a conservative mass-transfer. This distance is essentially a lower threshold for the distance of the source, due to our choice of fixing the donor mass to its minimum value. Furthermore, since switching on and off the magnetic braking in our conservative model (Equation \ref{eq:mdot}) gives two alternative expected luminosities, we distinguish between two lower limits for the distance, i.e. the "GR-only" distance $d_{m,GR}$ and the "MB-included" distance $d_{m,GR+MB}$. The final results are presented in Table \ref{tab:results}. In the same Table, the results for XTE J0929-314, already presented by \cite{Marino2017}, are included in order to give a general picture of all the AMXPs to which this method has been applied. For ease of discussion, each source is labeled with an arbitrary number, as shown in Table \ref{tab:results}, and in the following it will be indicated using this number. 

\begin{table*}
\centering
\caption{\textbf{Results}}
\begin{tabular}{l l  l l l l  l l  l   l l}
\hline 
\hline
& & M$_{min}$ & \textbf{L$_{exp, GR}$} & \textbf{L$_{exp,GR+MB}$} & \textbf{F$_{av}\times4\pi $d$^2_{kpc}$} & \textbf{d} & & \textbf{d$_{m,GR}$}	& \textbf{d$_{m,GR+MB}$}\\
\# & {\textbf{Source}} & M$_\odot$ & {\tiny(10$^{35}$ erg s$^{-1}$)} &  {\tiny (10$^{35}$ erg s$^{-1}$)} &{\tiny (10$^{33}$ erg s$^{-1}$)}& (kpc) & {\small \textbf{Method}} & (kpc) & (kpc) \\
\hline
S1 & IGR J18245-2452 & 0.21 & 0.21 & 1.40 & 1.32(13) & 5.5 $^{[4]}$ & GC & 3.97(19) & 10.2(5)\\
S2 & Swift J1749.4–2807 & 0.60 & 1.70 & 12.0 & 0.23(2) & <8$ ^{[13]}$ & B & 27.1(1.4) & 71.0(4.0) \\
S3 & IGR J17591-2342 & 0.43 & 2.20 & 13.0 & 5.2(5) & <7 $^{[14]}$ & M & 6.7(0.3) & 16.1(0.8) \\  
S4 & SAX J1748.9–2021 & 0.12 & 0.11 & 0.44 & 45.0(4.) & 8.5 $^{[8]}$ & GC & 0.50(2) & 0.99(5)\\
S5 & XTE J1814–338 & 0.20 & 3.20 & 5.40 & 0.49(5) & <9.6 $^{[10]}$ & B & 25.7(1.3) & 33.2(1.7) \\
S6 & IGR J17498–2921 & 0.20 & 1.80 & 3.70 & 0.61(6) & 7.6 $^{[7]}$ & B & 17.4(0.8) & 24.8(1.2)\\
S7 & IGR J17511–3057 & 0.16 & 2.50 & 3.80 & 2.30(20) & <7 $^{[1]}$ & B & 10.4(5) & 12.8(6) \\
S8 & IGR J00291+5934 & 0.04 & 0.35 & 0.41 & 0.98(9) & <4.7 $^{[2]}$ & B & 6.0(3) & 6.4(3) \\
S9 & SAX J1808.4–3658 & 0.008 & 0.95 & 1.01 & 14.0(1.4) & 3.5 $^{[3]}$ & B & 2.63(13) & 2.77(14) \\
S10 & IGR J1737.9-3747 & 0.07 & 2.30 & 2.50 & 2.20(20) & <4 $^{[14]}$  & M & 10.1(4) & 10.6(4) \\
S11 & HETE J1900.1-2455 & 0.018 & 0.32 & 0.33 & 790.0(80) & <5 $^{[5]}$ & B & 0.20(1) & 0.20(1) \\
S12 & NGC6440 X–2 & 0.0005 & 0.17 & 0.18 & 0.144(14) & 8.5 $^{[8]}$ & GC & 11.0(4) & 11.0(4) \\
S13 & Swift J1756.9–2508 & 0.008 & 0.19 & 0.20 & 0.41(4) & <14 $^{[14]}$  & M & 6.9(7) & 7.0(7)\\
S14 & IGR J16597–3704 & 0.007 & 0.25 & 0.25 & 2.5(2) & 9.1 $^{[11]}$ & GC & 3.14(16) & 3.15(16) \\
S15 & MAXI J0911–655 & 0.027 & 4.00 & 4.05 & 8.9(9) & 9.5 $^{[12]}$ & GC & 6.7(7) & 6.7(7) \\
S16 & XTE J0929–314 & 0.01 & 0.55 & 0.56 & 0.80(8) & U & - & 8.3(6) & 8.4(6)\\
S17 & XTE J1751–305 & 0.016 & 1.50 & 1.50 & 1.16(12) & <8.5 $^{[9]}$ & T & 11.4(4)& 11.5(4) \\
S18 & XTE J1807–294 & 0.0077 & 0.41 & 0.42 & 5.7(6) & U & - & 2.70(13) & 2.71(13) \\
S19 & IGR J17062-6143 & 0.0060 & 0.29 & 0.29 & 10.0(1.0) & 7.3 $^{[6]}$  & B & 1.66(8) & 1.66(8) \\
\hline
\hline
\end{tabular}
\caption*{List of the obtained expected and observed luminosities for the selected sample of AMXPs. The minimum masses for the companion star $M_{min}$ reported in this Table have been inferred from the mass function of the source fixing $i$ to 60$^\circ$, with the exception of the eclipsing Swift J1749.4-2807, for which $i$ was fixed to 90$^\circ$. In the Table $L_{exp,GR}$ and $L_{exp,GR+MB}$ are the expected luminosity evaluated for a mass-transfer driven by GR and GR+MB respectively, $F_{av}$ is the bolometric (0.1-300 keV) unabsorbed flux averaged over 20 years, $d$ the distance known for the system, $d_{m,GR}$ and $d_{m,GR+MB}$ the distances evaluated using $L_{exp,GR}$ and $L_{exp,GR+MB}$ respectively. The label "U" indicates that the distance of the source is unknown at the moment. The cited values for $d$ are upper limits on the distance, therefore we reported the value found on the refereed paper plus its upper error. In the "Method" column we indicate the method used to determine the upper limit on $d$ reported here. \\
Methods: "B": study of type-I X-ray bursts. "GC": association to a Globular Cluster of known distance. "M": study of the extinction maps by \cite{Marshall2006}. "T": theoretical argument. \\
References: [1]=\cite{Altamirano2010_17511}, [2]=\cite{DeFalco2017}, [3]=\cite{Galloway2006_1808}, [4]=\cite{Harris1996}, [5]=\cite{Kawai2005}, [6]=\cite{Keek2017_j17062}, [7]=\cite{Linares2011}, [8]=\cite{Ortolani1994}, [9]=\cite{Papitto2008}, [10]=\cite{Strohmayer2003}, [11]=\cite{Valenti2007}, [12]=\cite{Watkins2015}, [13]=\cite{Wijnands2009}., [14]=This paper.}
\label{tab:results}
\end{table*}

\section{Results}
\label{sec:disc}
According to the general picture emerging from Table \ref{tab:results}, we discuss the sources distinguishing first between sources with unknown distance values and sources with a reported distance estimate (or at least an upper limit). For the latter sources, i.e. 14 over 19 sources in the sample, the comparison with the estimated distance lower limits is obviously easier, while in the discussion for the former sources the soundness of our distance limits will be considered. 
We checked if any of the available distance values could be updated using the public results of the Global Astrometric Interferometer for Astrophysics (\emph{GAIA}) \citet{Gaia.etal:2016a}, in the catalogue GAIA DR2 \citep{GaiaCollab2018}, excluding the Globular Clusters sources, since their distances are know with sufficient precision. For only 6 sources, i.e. S5, S7, S11, S17, S18 and S19 GAIA found counterparts within 2 arcsec error boxes. The distance ranges found for these optical counterparts are wide and always compatible with the distances reported in the literature (and in this paper) and the association of these sources with the optical systems in the \emph{GAIA} catalogue is not certain. We therefore decided not to include \emph{GAIA} results in this work. \\ 
Among the 14 sources with a known distance, in 6 cases, i.e. SAX J1748.9-2021, SAX J1808.4-3658, HETE J1900.1-2455, MAXI J0911-655, XTE J1807-294 and IGR J17062-6143, our method gave a lower limit for the distance which is lower than the measured distances. The method in these cases must be considered inconclusive, since changing the assumptions made, e.g. on the secondary mass, might fix the discrepancy between the distance values and indicate a compatibility with a conservative scenario.


Not surprisingly, these sources have emitted huge amounts of energy in the time considered. Two of these five sources have indeed undergone several outbursts in the last 20 years and three of them, i.e. S11, S15 and S19, have been persistently in outburst for years (see Table \ref{tab:outbursts}), even if, in the latter case, at a faint luminosity.  \\
Except for one source, i.e. S1, the remaining 7 sources with known distances have estimated values smaller than both the $d_{m,GR}$ and $d_{m,GR+MB}$ limits, indicating a likely non-conservative mass-transfer scenario. 
S1 is the only source where $d_{m,GR}$ $ < \ d \ < $ $d_{m,GR+MB}$. While this result might be considered inconclusive, it is indeed remarkable that the relatively long orbital period of the system, i.e. 11.03 hr \citep{Papitto2013}, seems to discourage a GR-only model, making the higher distance-limit (and the non-conservative scenario) more realistic. \\  
The limits on the distance discussed so far rely strongly on the choice to fix the time interval $T$ to 20 years. Taking into account smaller time intervals would also decrease the limits on the distance. In the following we release this assumption and we assume $T$ equal to the time elapsed since the first outburst of each source in the last 20 years\footnote{This time usually coincides with the time of discovery of pulsations for the source, displayed in Table \ref{tab:amxps}, with the exception of S10 and S19.} (see Table \ref{tab:outbursts}), in order to test the robustness of these results. The new choice for $T$ leaves almost unchanged the situation for S2, S5, S6 and S17, because even the new limits for d (i.e. 17, 22, 10 and 10 kpc respectively)  exceed the current distance estimate. On the contrary, considering a shorter $T$ changes substantially the results obtained for S1, S7, S8 and S12, which are now equal or even lower than the measured d, i.e. at 5, 7, 5, and 7 kpc respectively. This result reflects also the fact that these sources, especially S1, have been discovered very recently. 
Therefore we may conclude that we have a strong evidence for non-conservative mass-transfer for S2, S5, S6 and S17, while this evidence is more troublesome for the other four sources in the group.  \\
As stated before, for five sources in our sample we do not have available distance estimates. In this work we attempt to find at least an upper limit on these distance values using the 3D extinction maps of radiation in the $K_S$ band for our Galaxy computed by \cite{Marshall2006}\footnote{\url{http://vizier.cfa.harvard.edu/viz-bin/VizieR?-source=J/A+A/453/635}}, following the same steps by \cite{Gambino2016}. Even if on the basis of its location, i.e. very close to the Galactic Center, an upper limit of 8.5 kpc was suggested for XTE J1751-305 \citep{Papitto2008}, no direct measurements of its distance are available so far, therefore we included the source in the sample and tried to evaluate its distance too.
For a specific direction in the Galaxy, e.g. defined by the coordinates of a source, these maps give the evolution of the extinction in the $K_S$ band, $A_{K_S}$, as a function of the distance. In order to evaluate $A_{K_S}$, we used the equations:
\begin{equation}
    N_H=(2.21\pm0.09)\times10^{21}A_V ,
\end{equation}
\begin{equation}
    A_{K_S}=(0.062\pm0.005)A_V \ \textrm{mag} ,
\end{equation}
by \cite{Guver2009} and \cite{Nishiyama2008} respectively, where $A_V$ is the extinction in the visual band. The references for the $N_H$ values employed for each source are reported in the "References Spectral Parameters" column in Table \ref{tab:outbursts}. \\
From the study of the extinction maps, we derive distances of 5$\pm$2 kpc for S3, 12$\pm$2 kpc for S13 and an upper limit of 4 kpc for the distance of S10. Our estimate for the distance of S3 is barely compatible with the lower limit posed by \cite{Russell2018} of 6 kpc. \\ Results for S17 are inconclusive, since the only constraints on the distance found is a lower limit of 2.7 kpc. S16 and S18 have relatively high Galactic latitude, in directions poorly mapped by \cite{Marshall2006}, therefore even with this technique it is not possible to obtain constraining limits on the distances. The discussion about the non-conservativity of the mass-transfer has to shift towards how reasonable is our threshold distance for these two sources. Taking into account the new distance upper limits, S10 results compatible with a non-conservative scenario according to both $d_{m,GR}$ and $d_{m,GR+MB}$, while the limits for S13 are below the known distance for the source, making the method inconclusive. The situation of S3, for which $d_{m,GR} \ < \ d \ < \ d_{m,GR+MB}$, recalls what found for S1. By the way, since the long orbital period of the source, i.e. 8.82 hr \citep{Sanna2018}, strongly encourages a MB contribution in the dynamics of the system, the non-conservative case seems to be better-founded than the conservative case. The distance limits for S18 are reasonable and they have to be considered inconclusive for our purposes. Also S16 has a realistic distance threshold, although its relatively high Galactic latitude would place it in an empty region of the Galaxy, suggesting a non-conservative mass-transfer scenario \citep{Marino2017}. We applied the test of changing the choice for $T$ to S3 and S10. While S3 was discovered only last year and it is clearly not suitable for such test, S10 has upper limits with the new $T$ of $d_{m,GR}$=8.4$\pm$0.5 kpc and $d_{m,GR+MB}$=8.9$\pm$0.5 kpc, which are still suggestive of a non-conservative scenario. We therefore include S10 in the sample of our strong evidences sources.\\
It is noteworthy that if no activity is observed in the next few years, the limits for $d$ found for the 19 sources analyzed in this paper will drift to higher values and some of the weak evidences for non-conservative mass-transfer might be strenghtened.
\section{Discussion}
\begin{figure*}
\centering
\includegraphics[height=10 cm, width=16 cm]{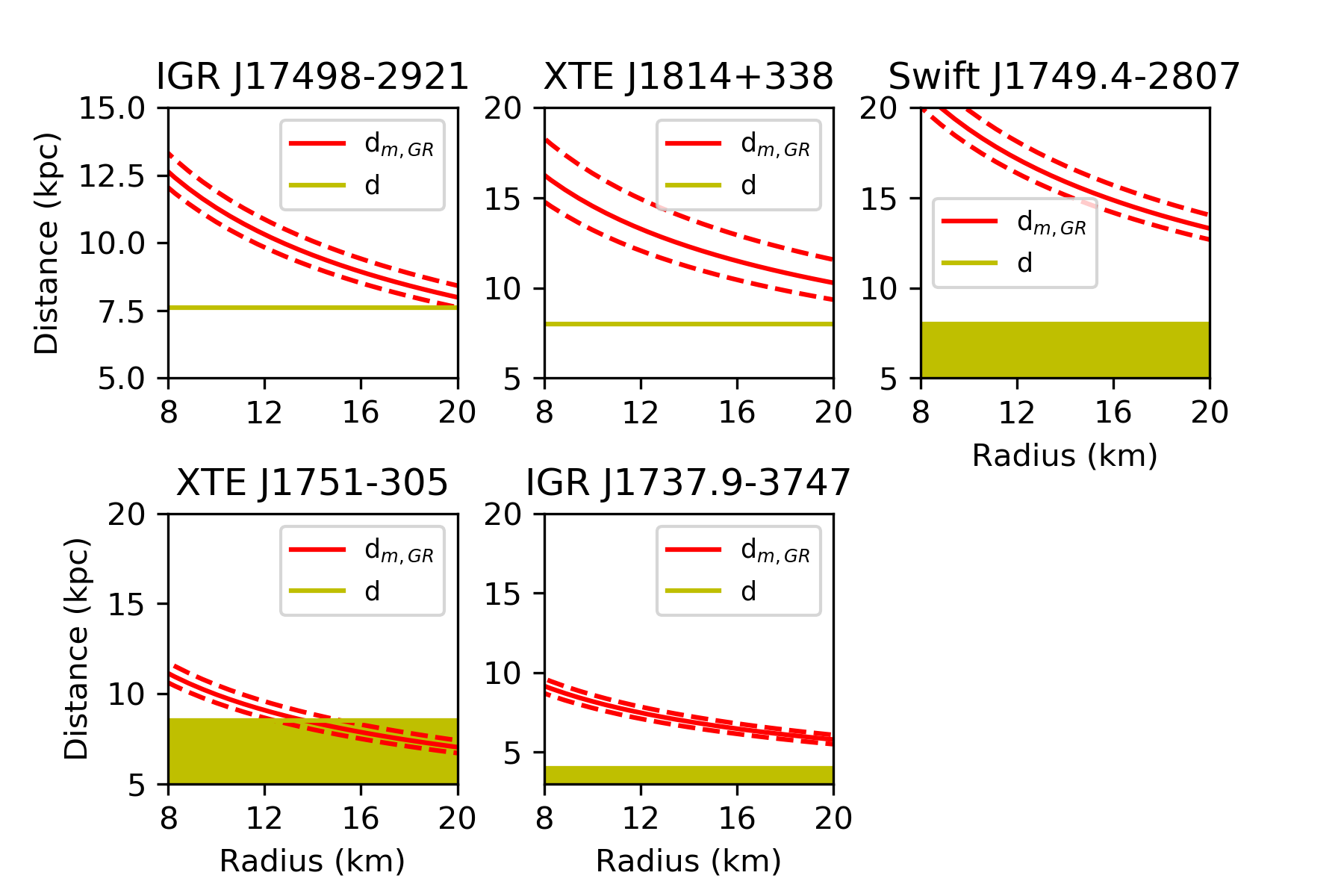}
\caption{Distance-Radius curves (in red) with errors (dashed lines) compared with the distance value reported in literature (in yellow) for the four sources with stronger evidences for non-conservative mass-transfer.}
\label{fig:Radius}
\end{figure*}

The method described and applied in this paper strongly indicates a non-conservative mass-transfer for six sources (including the results for XTE J0929-314), and weakly for other five sources. \\
We discuss here the possibility that any of the assumptions made in this work may have biased our results for the strong evidences. In the following, $T$ was considered the time elapsed since the discovery of the source. First of all, we checked  decreasing the value assumed for the mass of the NS to 
1.1 M$_\odot$ \citep[lower than the lowest NS mass ever measured, i.e. 1.174$\pm$0.001, ][]{Martinez2015} and increasing the radius of the neutron star above 12 km \citep[i.e. above the values expected according to most proposed Equations of state, ][]{Ozel2016} might give different results. Fig. \ref{fig:Radius} shows the distance as a function of $R^{-1/2}$, using for $F_{av}\times4\pi d^2$ and $\dot{M}_{min}$ the values previously evaluated for the five strong evidence sources. For four sources out of five, an unrealistic value of  $R$ > 20 km would be needed to match the luminosity expected under the assumption of conservative mass transfer. In XTE J1751-305, on the contrary, the compatibility is restored for values of $R$ >12 km. We therefore discard this source from the strong evidences sources.


Another assumption which might be relaxed concerns the inclination, fixed to $60^\circ$ in all the sources but the only eclipsing one, i.e. Swift J1749.4-2807. However, allowing for a 90$^\circ$ inclination does not give substantial changes in the obtained values and the scenario proposed here is unchanged. \\
Even in the extreme scenario of 90$^\circ$-inclination systems (for which no eclipsing activity was ever observed), hosting bizarre neutron stars with high radii and the smallest mass value ever observed we are not able to find an accordance between data and the predictions of a conservative mass-transfer model for these sources.\\ 
Even if we consider wrong the distance measured for these sources, we still have to admit high inclinations and low-mass companions to obtain distances within the size of our Galaxy. \\ Finally, our choice for $T$ in this discussion was the most conservative possible and replacing it with 20 years may only reinforce these evidences.
Therefore, the Occam's razor leads us to consider these results as a strong evidence for non-conservative mass-transfer for IGR J17498-2921, XTE J1814+338, Swift J1749.4-2807, IGR J1737.9-3747. 
As stated in Section \ref{sec:intro}, a non-conservative mass-transfer has been proposed in three other AMXPs: XTE J0929-314, SAX J1808.4-3658 and SAX J1748.9-2021. The latter two sources have been analyzed as well in this paper, giving no evidence for a non-conservative mass-transfer. This should not be considered a contradiction, because the possibility of a non-conservative mass-transfer is not excluded by this method; allowing for higher companion masses might increase the expected luminosity and shift $d_{m,GR}$ ($d_{m,GR+MB}$) below the distance estimate for the system. It is impossible, however, with the current available information about the inclination and/or the secondary mass of these sources to discriminate between the conservative and non-conservative scenarios. Furthermore, as recently demonstrated by \cite{Van2018}, our prescription for the MB term might not be the most adequate and may significantly underestimate the real contribution, i.e. by an order of magnitude, to the average theoretical mass-transfer. Indeed, the model used here neglects some physical aspects, e.g. the chemical or thermal evolution of the companion star, as well as any calculation of the convective and radiative zones of it, therefore these results have to be considered with a reasonable dose of caution. These arguments might explain why S4 and S9 do not give evidence for a non-conservative mass-transfer although a non-conservative evolution has been hypothesized \citep{DiSalvo2008,Sanna2016,Sanna2017_1808} with an independent argument, in order to justify the large value of the orbital period derivative measured for both systems\footnote{A short-term variability induced by tidal dissipation and magnetic activity in the companion, which is required to be at least partially non-degenerate, convective, and magnetically active \citep{Applegate1994,Hartman2008}, has been alternatively invoked to explain the observed orbital period expansion, although this possibility appears quite unlikely \citep{Sanna2017}}. \\

It is still unclear the physical mechanism inhibiting a fully conservative mass-transfer in these systems. A model which would explain this phenomenon was proposed by \citet{Burderi2001} and it is known as the \emph{radio ejection} model. It predicts that a non-conservative mass-transfer may arise in LMXBs hosting rapidly rotating pulsars when the radiative pressure of the pulsar, emitting as a magnetic-dipole rotator, overcomes the ram pressure of the accreting matter, throwing it away from the system. \\
Furthermore, a recent work by \citet{Ziolkowski2018} shows with a similar argument how also LMXBs hosting BHs might undergo a secular evolution driven by non-conservative mass-transfer. In these systems, radio ejection is clearly out of the picture; in that case, matter outflows in the form of jets and winds have been invoked to explain the phenomenon. \\

\section{Conclusions}
This work aims at a discussion of the possibility of a non-conservative mass-transfer for almost all the AMXPs discovered so far by comparing their expected luminosity, calculated on the conservative evolution hypothesis, and the observed X-ray flux averaged over the last 20 years (or the time elapsed since their discovery). Recently a non-conservative mass-transfer has been claimed for three AMXP; including the four strong evidence cases presented here the count would increase to seven, suggesting that this physical phenomenon might be not rare at all in the family of the AMXPs. Furthermore, if we include also the sources with weak evidence of non-conservative mass-transfer and the two AMXPs which displayed only faint outbursts (for which conservative mass-transfer rates seem unlikely), more than half of the AMXPs would require a non-conservative scenario. It is also interesting to notice how, despite more than 20 years of continuous monitoring by several all-sky 
monitors, the number of transient LMXBs showing millisecond pulsations keeps rising \citep[the latest discovered is IGR J17591-2342,][]{Sanna2018}. This argument might be taken as a hint that the recurrence times could be even longer than 15 years, enlightening the inadequacy of conservative mass-transfer scenarios for many members of this class.\\ 
The radio ejection model, developed several years before these pieces of evidence for non-conservative mass-transfer were found, might be able to explain what induces a non-conservative regime in the case of accreting fast pulsars and, if confirmed, it could be considered as a key feature in the complex phenomenology of this class of sources.

\begin{acknowledgements}
The authors are grateful to the anonymous referee for a thoughtful report that helped improve the results and the clarity of this work. The authors acknowledge financial contribution from the agreements ASI-INAF I/037/12/0 and n.2017-14-H.0. This work is partially supported by the HERMES Project,
financed by the Italian Space Agency (ASI) Agreement n. 2016/13 U.O. AP acknowledges funding from the European Union’s Horizon 2020 Framework Programme for Research and Innovation under the Marie Sklodowska-Curie Individual Fellowship grant agreement 660657-TMSP-H2020-MSCAIF-2014. This work has made use of data from the European Space Agency (ESA) mission Gaia (\url{https://www.cosmos.esa.int/gaia}), processed by the Gaia Data Processing and Analysis Consortium (DPAC, \url{https://www.cosmos.esa.int/web/gaia/dpac/consortium}). Funding for the DPAC has been provided by national institutions, in particular the institutions participating in the Gaia Multilateral Agreement. AM acknowledges funding from FSE (Fondo Sociale Europeo) Sicilia 2020.
\end{acknowledgements}

\bibliographystyle{aa}
\bibliography{biblio}
\newpage 

\begin{appendix} 
\label{appendix}
\section{Spectral Analysis of XTE J1814+338}
\begin{figure}
\centering
\includegraphics[height=5 cm, width=7 cm]{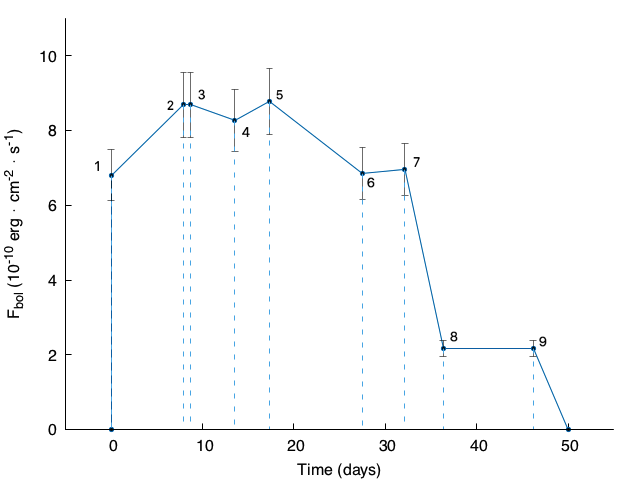}\llap{\raisebox{2.8cm}{\includegraphics[height=1.4 cm]{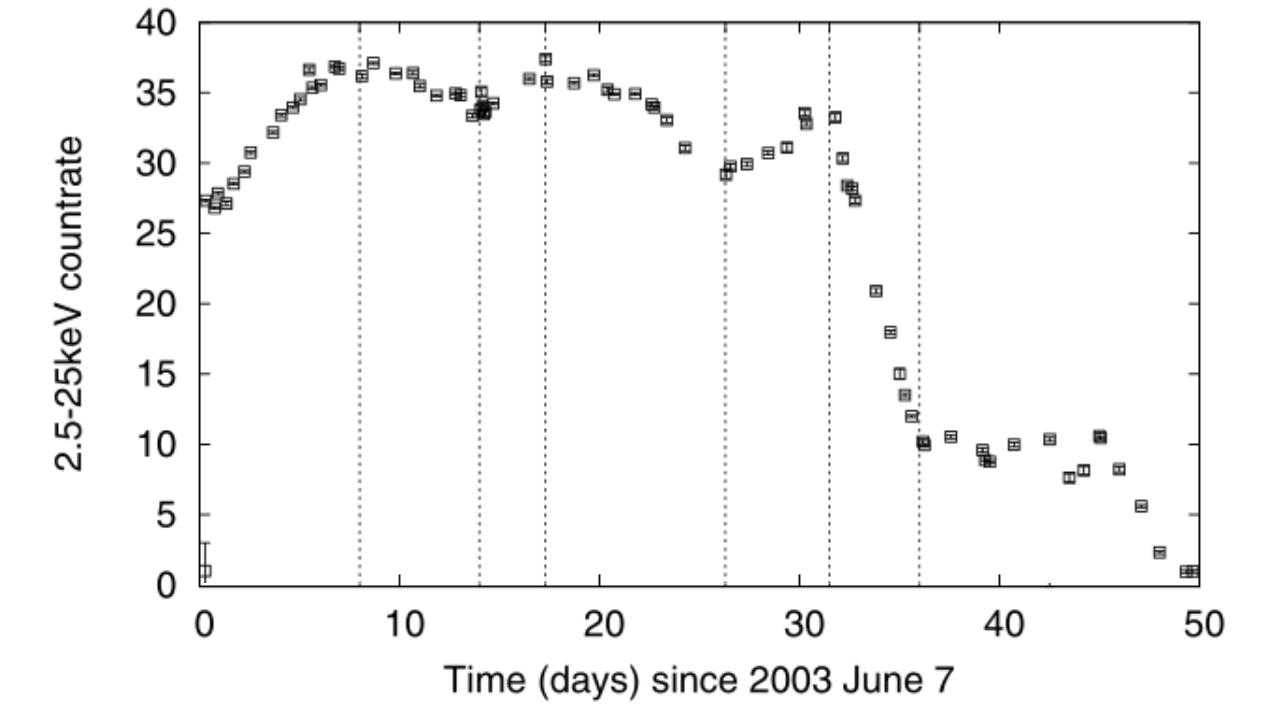}}}
\caption{ Data points used to describe geometrically the light curve of the 2003 outburst of XTE J1814-338. In this representation, the light curve is described as the piecewise function connecting the data points and its area is obtained by dividing it in trapezes, whose vertical lines are the dashed lines in this figure, and summing their areas; the light curve is shown for comparison in the box above \citep{Papitto2007}.}
\label{fig:XTE}
\end{figure}

Unlike any other source in the sample, spectral information about XTE J1814+338 during its one and only outburst in 2003 are not reported in literature. Furthermore the \emph{RXTE/PCA} (Proportional Counter Array) light curve of the 53-days long outburst does not have a simple shape comparable to any geometrical figure (see \citet{Papitto2007}, Figure 1): after a smooth rise lasting for $\sim$ 5 d, the emitted flux stabilized to an order of magnitude of 10$^{-10}$ erg cm$^{-2}$ s$^{-1}$ (2.5-25 keV) for 33 d, with a peak flux of 5$\times$ 10$^{-10}$ erg cm$^{-2}$ s$^{-1}$ in the same energy range and fluctuations around 20\%, until it decayed abruptly to one fourth of the peak flux. In order to calculate the fluence of the source we performed a simple spectral analysis of the system during the outburst.\\
In order to calculate the area subtended by the curve we described the light curve with a piecewise function, as shown in Figure \ref{fig:XTE}, and then we divided the whole area in 8 trapezes. The area of these trapezes is easy to calculate once we have the information about the flux at the beginning and at the end of the corresponding time segment. Therefore we extracted the RXTE spectra from the RXTE standard products (in particular source and background spectra and the response file) of the source in 7 strategic points, in order to get the bolometric (0.1-300 keV) flux. It is important to notice that, according to the light curve shape, we considered the flux constant over two time windows, i.e. between points 2-3 and 8-9 in Figure \ref{fig:XTE}. \\ Standard data products are not meant to be used for a detailed analysis of the source, but they are suitable for our purpose to have rough information about the bolometric flux evolution from the source. The spectrum of each observation was analysed using the spectral package Heasarc \textsc{Xspec} v. 12.9.1, and fitted to a simple power law, described by the \textsc{powerlaw} model, multiplied by \textsc{tbabs} to take into account the interstellar absorption, with \textsc{vern} cross-sections \citep{Verner1996} and \textsc{wilms} abundances \citep{Wilms2000}. Observations log and spectral fit results are reported in Table \ref{tab:fit}. 

\begin{table*}
\centering
\caption{\textbf{List of the observations used in this work and fits results}}
\begin{tabular}{l  l  l  l  l  l  l }
\hline 
\hline
& {\textbf{ObsID}} & \textbf{Date} & \textbf{$\Gamma$ index} & \textbf{N$_H$} & \textbf{F $_{bol}$ \textbf(0.1-300 keV)} & \textbf{$\chi^2_{\nu}$}	\\
&	&  (mm-dd-yyyy) & & {($\times$ 10$^{22}$ cm$^{-2}$)} & {($\times$ 10 $^{-10}$ erg $\times$cm$^{-2}$ $\times$ s$^{-1}$ )}& {(d.o.f.)}\\
\hline
\hline
{1} & {80145-02-01-01} & {06-07-2003}& {1.600$\pm$0.020} & {$<$0.48}&{6.80$\pm$0.68} & {1.45(47)}\\
{2} & {80418-01-02-06} & {06-15-2003}&{1.706$\pm$0.012}&{0.89$\pm$0.16}&{8.69$\pm$0.87}&{0.99(46)}\\
{3*} \\
{4} & {80418-01-03-12} & {06-20-2003}&{1.659$\pm$0.015}&{0.40$\pm$0.20}&{8.27$\pm$0.83}&{0.90(46)}\\
{5} & {80418-01-03-08} & {06-24-2003}&{1.670$\pm$0.020}&{0.45$\pm$0.30}&{8.78$\pm$0.88}&{0.86(46)}\\
{6} & {80418-01-05-00} & {07-04-2003}&{1.700$\pm$0.030}&{0.50$\pm$0.40}&{6.85$\pm$0.69}&{0.71(46)}\\
{7} & {80418-01-05-09} & {07-09-2003}&{1.760$\pm$0.013}&{0.300$^{\dagger}$}&{6.96$\pm$0.70}&{0.93(48)}\\
{8} & {80418-01-06-02} & {07-13-2003}&{1.840$\pm$0.020}&{0.300}$^{\dagger}$&{2.16$\pm$0.22}&{0.97(47)}\\
{9*}\\
\hline
\hline
\end{tabular}
\caption*{The data have been all analyzed using a simple \textsc{tbabs}$\times$\textsc{powerlaw} model. The quoted errors for $\Gamma$ and $N_{H}$ were calculated with the \textsc{error} command for the resulting simulated posterior distribution and reflect 90\% confidence intervals, while a standard uncertainty of 10$\%$ has been attributed to each of the bolometric flux values. \\
*: Under the assumption that the flux was almost the same between the previous data point and this one, an assumption based on the light curve shape over these time intervals, the bolometric flux at this time was considered equal to the flux reported in the preceding row;\\
$\dagger$: the parameter was kept frozen to the reported value for the stability of the fit.}
\label{tab:fit}
\end{table*}

\end{appendix}
\end{document}